# Double Integral Enhanced Zeroing Neural Network Optimized with ALSOA fostered Lung Cancer Classification using CT Images

V S Priya Sumitha [1],   V. Keerthika [2]   A. Geetha [3]

priyasumitha123@gmail.com[1]

keerthika.v@alliance.edu.in[2]

geetha.a@alliance.edu.in[3]

KG  College of Arts and Science[1]

Alliance college of Engineering and Design,[2,3]
Bangalore

## Abstract

Lung cancer is one of the deadliest diseases and the leading cause of illness and death. Since lung cancer cannot predicted at premature stage, it able to only be discovered more broadly once it has spread to other lung parts. The risk grows when radiologists and other specialists determine whether lung cancer is current. Owing to significance of determining type of treatment and its depth based on severity of the illness, critical to develop smart and automatic cancer prediction scheme is precise, at which stage of cancer. In this paper, Double Integral Enhanced Zeroing Neural Network Optimized with ALSOA fostered Lung Cancer Classification using CT Images (LCC-DIEZNN-ALSO-CTI) is proposed. Initially, input CT image is amassed from lung cancer dataset. The input CT image is pre-processing via Unscented Trainable Kalman Filtering (UTKF) technique. In pre-processing stage unwanted noise are removed from CT images. Afterwards, grayscale statistic features and Haralick texture features extracted by Adaptive and Concise Empirical Wavelet Transform (ACEWT). Then, Double Integral Enhanced Zeroing Neural Network (DIEZNN) is utilized to categorize the image into Cancer, Non-cancer. Artificial Lizard Search Optimization Algorithm (ALSOA) is used to optimize DIEZNN classifier, precisely classifies lung cancer images. The proposed model is implemented on MATLAB. The performance of the proposed method is analyzed through existing techniques. The proposed method attains 18.32%, 27.20%, and 34.32% higher accuracy analyzed with existing method likes Deep Learning Assisted Predict of Lung Cancer on Computed

Tomography Images Utilizing AHHMM (LCC-AHHMM-CT), Convolutional neural networks based pulmonary nodule malignancy assessment in pipeline for classifying lung cancer (LCC-ICNN-CT), Automated Decision Support Scheme for Lung Cancer Identification with Categorization (LCC-RFCN-MLRPN-CT) methods respectively.

## 1. Introduction

Lung cancer or lung carcinoma is term for unchecked cells development lung tissue caused by a malignant tumor. Treatment should be started right away to stop it from using metastases to spread to other body organs [1]. Major two types of lung carcinomas such as lung carcinoma of minor cell, lung carcinoma of non-minor cell represent the bulk of lung malignancies' first stages. In 88% of instances, long-term smoking is primary contributor of lung cancer. In those who have never smoked, instances are caused by factors such as asthma, asbestos, radon gas in about 15 to 20% of cases [2]. The traditional approaches for identifying lung cancer are Computed Tomography (CT) and X-rays. Usually, a biopsy is performed in conjunction with a CT scan or bronchoscope to confirm the diagnosis. Since lung cancer primary cause of cancer-related demise, an innovative then all-encompassing approach to lung cancer early diagnosis must be devised [3]. Recent advancements in deep learning and deep neural network technologies have enhanced picture recognition. Using Deep Neural Networks, it is possible to scan patterns in photos and determine whether they are recognized or not. Additionally, certain patterns should be sought for when evaluating an image. Predetermined datasets have been used by the network to detect, learn, and categorize images in order to train neural networks [4]. Deep Neural Network (DNN) is becoming more and more well-liked since it makes categorizing and identifying picture patterns straightforward. Despite major technological breakthroughs radiation preparation, imaging and treatment outcomes remain very subpar. Increasing radiation could provide a way to strengthen the therapeutic effects [5].

Despite existing technological advancements in radiation imaging and preparation, the treatment outcomes are still relatively poor. The existing technique does not offer the enough accuracy and RoC while performing the task. These drawbacks in the existing approaches inspired to do this work.

Major contributions of this research is abridged below,

- ❖ Double Integral Enhanced Zeroing Neural Network Optimized with ALSOA fostered Lung Cancer Classification utilizing CT Image is proposed.
- ❖ In this research, noises are removed by Unscented Trainable Kalman Filtering and Adaptive with Concise Empirical Wavelet Transform used for feature extraction.
- ❖ Then introduces Double Integral Enhanced Zeroing Neural Network (DIEZNN) to classify the CT image into cancer and non-cancer and DIEZNN is optimized with Artificial Lizard Search Optimization Algorithm (ALSOA).
- ❖ The performance metrics like accuracy and RoC is analyzed to compare the proposed method performance with other existing techniques such as LCC-AHHMM-CT, LCC-ICNN-CT, and LCC-RFCN-MLRPN-CT.

Continual manuscript is structured as below: part 2 reviews literature review, proposed method demonstrated in part 3, results are proved in part 4, conclusion is depicted in part 5.

## 2. Literature Review

Several studies on deep learning based lung cancer classification were presented by literature. The current studies were covered below,

In 2020, Yu, H., et.al, [6] have presented deep learning assisted predict of lung cancer on Computed Tomography Images Utilizing AHHMM. An Adaptive Hierarchical Heuristic Mathematical Model (AHHMM), a deep learning technique, has been suggested. To evaluate DL based on the preceding therapy plan in establishment of automatic radiation adaption procedures for Non-Small Cell Lung Cancers (NSCLC) seek to optimize local tumor regulation at lesser rates of grade 2 RP2 radiation pneumonitis. It provides higher accuracy and lower error rate.

In 2020, Bonavita, I., et.al, [7] have presented convolutional neural network based pulmonary nodule malignancy assessment in pipeline for classifying lung cancer. The nodule malignancy was analyzed using 3D convolution neural networks and results were integrated into the existing computerized endwise workflow for lung cancer analysis. Different LIDC data set parts were used for training and testing. While adding nodules to the lung cancer pipeline raised the likelihood of lung cancer by 14.7% F1-weighted score. It provides higher accuracy and lower error ratio.

In 2020, Masood, A., et.al, [8] have presented Enhanced RFCN coupled with Multilayer Fusion RPN: Automated Decision Support System for Lung Cancer Identification with Categorization. An enhanced multi-dimensional region-based fully convolutional network base automatic decision support scheme was presented for lung nodule recognition, categorization. A novel multi-Layer fusion region proposal network was used as backbone of the image classifier for feature extraction is studied. According to experiment results, a promising detection performance obtains high sensitivity and low accuracy when analyzed to state-of-the-art nodule identification.

In 2021, Sori, W.J., et.al, [9] have presented deep learning-based DFD-Net: detecting lung cancer from denoised CT scan images. It was made up entirely of denoising and detecting components. First, noise during the preprocessing step was removed using residual learning Denoising model. Next, two-path CNN was used to discover lung cancer utilizing denoised image produced via DR-Net as an input. The combined integration of local and global aspect was the main focus of the two routes. Each path has a distinct receptive field size to achieve, which helps to describe both local and global interdependence. It provides high accuracy and low error rate.

In 2022, Shakeel, P.M., et.al, [10] have presented improved deep neural networks and ensemble classifiers for automated lung cancer recognition from CT image. Here, lung cancer was predicted using a brand-new, enhanced image processing and machine learning technology. The goal of collecting images from NSCLC CT scan datasets was to identify lung cancer. The most useful features were then selected utilizing an ensemble classifier and HSOA with an intelligent generalize rough set method. It provides higher precision and higher F-score.

## 3. Proposed Methodology

In this section, Double Integral Enhanced Zeroing Neural Network Optimized with ALSOA fostered Lung Cancer Classification using CT Images (LCC-DIEZNN-ALSO-CTI) is deliberated. The block diagram of LCC-DIEZNN-ALSO-CTI Lung Cancer Classification System is shown in figure 1. It comprises data acquisition, pre-processing, feature extraction, classification and optimization. Then, detailed about depiction every stage given below,

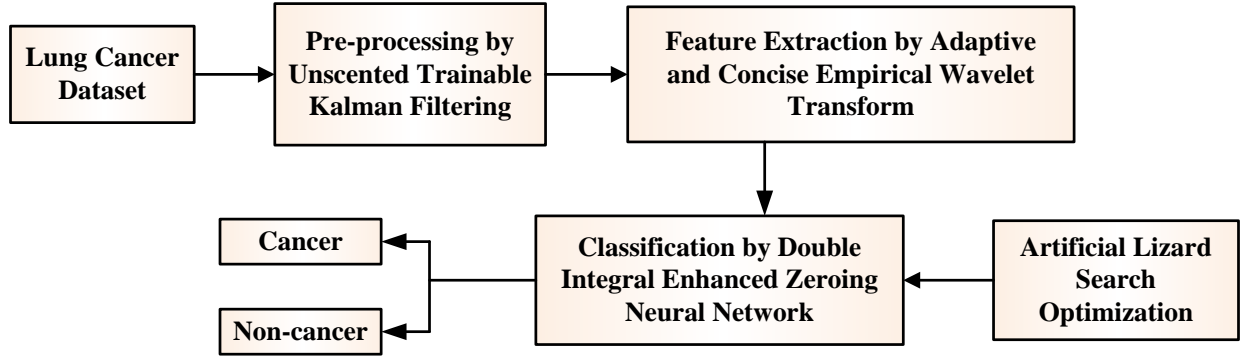

**Figure 1: LCC-DIEZNN-ALSO-CTI Lung Cancer Classification System**

### 3.1 Data Acquisition

Initially, the input CT image taken from lung cancer dataset [11]. The majority of the time, research analysis uses this standardized dataset. There are only 32 instances in the raw dataset. Through the use of data augmentation, the research's data samples are increased to 488. The data takes into account common variables like gender, age, and chest discomfort. In addition to these, complex symptoms including hereditary diseases, alcoholism, and smoking habits, among others, is also used as research data cases. All attributes are of the integer data type, and the domain range for each attribute is labeled on a scale of 1 to 10.

### 3.2 Pre-processing using Unscented Trainable Kalman Filtering

Pre-processing of input CT image is discussed here. Unscented Trainable Kalman Filtering (UTKF) [12] is used in pre-processing to remove noise and enhances the quality of CT images. When estimating the state of nonlinear systems, the Unscented Trainable Kalman Filter (UTKF), a Kalman Filter version, is used. It was created to address circumstances in which the conventional Extended Kalman Filter (EKF) might not function properly because of significant nonlinearity. Following the prediction process, sigma points are given in equation (1)

$$A^{(1)}(s+1) = \hat{a}(s+1) \tag{1}$$

Where, $A^{(1)}(s+1)$ indicates $1^{st}$ sigma point of forecast $a(s+1)$ and $\hat{a}(s+1)$ denotes the noise prediction vector. The unscented transition, which could transfer variables then not modify pixel values of CT images, employed to remove image noise in order to solve nonlinear function and reduce estimate error. The $l^{th}$ sigma point value is given in equation (2)

$$A^{(l)}(s+1) = \hat{a}(s+1) + \left[\sqrt{(l+\beta)\sum_a(s+1)}\right]l \quad (2)$$

Where, $A^{(l)}(s+1)$ represents $l^{th}$ sigma point of forecast $a(s+1)$, $l$ indicates the random value from 2 to $n+1$. $\beta$ is a pixel value of input CT image. $\sqrt{.}$ indicate the Cholesky factorization function. Following the creation of sigma points, the measurement procedure may be characterized as given in equation (3), (4)

$$\delta^{(l)}(s+1) = g(A^{(l)}(s+1)), k = 1,...,2m+1 \quad (3)$$

$$\hat{C}(s+1) = \sum_{l=1}^{2m+1} \eta^{(l)} \delta^{(l)}(s+1)$$

(4)

Where, $\delta^{(l)}(s+1)$ refers the sigma reference point; $g$ is a nonlinear vector and the value of $\eta^{(l)} = 1/2M$. $M$ indicates the number of noise present in CT images. The output of the filter is given in equation (5)

$$\sum_C(s+1) = \sum_{l=1}^{2m+1} \eta^{(l)} \left(\delta^{(l)}(s+1) - \hat{C}(s+1)\right) + R(s+1) \quad (5)$$

Where, the variance matrix for the measurement is refers as $\sum_C(s+1)$, and the noise matrix for the measurement method is $R(s+1)$. Here the unwanted noises are removed. Thus the pre-processing of input CT images is done by Unscented Trainable Kalman Filtering. The pre-processing output given to feature extraction stage.

### 3.3 Feature Extraction utilizing Adaptive and Concise Empirical Wavelet Transform

The feature extraction technique is discussed. Adaptive with Concise Empirical Wavelet Transform [13] is utilized to extract Radiomic features such as Grayscale statistic features (standard deviation, mean, kurtosis, and skewness) and Haralick Texture features (contrast, energy, entropy and homogeneity) from pre-processed CT images. The most important step in computing the modes is getting boundaries. Quantity and position of boundaries dictate the kind, degree of modes. The pixel value truly depicts the power and frequency distribution. Fourier transform of image $b(s)$ is $\hat{b}(e)$, and its average value is given in equation (6)

$$Q = \lim_{S \to \infty} \frac{1}{2S} \int_{-S}^{S} b(s)^2 ds \tag{6}$$

Where, $S$ indicates the time interval, $Q$ is an average pixel value of CT image. As long as the image's Fourier transform in the range [0, S] is given in equation (7)

$$\hat{b}_S(e) = \frac{1}{\sqrt{S}} \int_{0}^{S} b(s) e^{-u2\lambda es} ds \tag{7}$$

EWT performs effectively at component separation for noise-free images. But in practice, noise is challenging to eliminate. EWT encounters challenges since the pixels typically belong to a non-stationary state. Following that definition, the pixel density of pre-processed CT image is given in equation (8)

$$T_{bb}(e) = \lim_{S \to \infty} F\left[\left|\hat{b}_S(e)\right|^2\right]$$

(8)

Where, $F$ indicates the features present in the CT images. The features envelop vector value is given in equation (9)

$$\hat{O}_{bb}(\alpha) = \frac{1}{K-p} \sum_{u=1}^{K-p} |b_m(s_u)|^2 \cdot |b_m(s_u + \alpha)|^2 \tag{9}$$

Where, $p$ represents the extract features of the CT images. $\alpha$ denotes the delay factor and $K$ is an overall features present in the image. $S_u$ is a time taken for extract $u$ number of features and $b_m$ is a coefficient vector. By this, the Grayscale statistic features, Haralick Texture features extracted using ACEWT. Grayscale features like standard deviation, mean, kurtosis, skewness are discussed below,

Equation (10) is used to calculate **mean** of Grayscale statistic features.

$$\chi = \frac{\sqrt{\sum_{k=1}^{m} p}}{Z} \tag{10}$$

Where, $\chi$ is the mean, $Z$ indicates the total pixel value.

**The standard deviation** of the CT image feature can be expressed in equation (11)

$$SD = \sqrt{\frac{1}{A}\sum_{a=1}^{A}(B(a)-\chi)^2} \qquad (11)$$

The **Kurtosis** of CT image feature may be expressed by using equation (12)

$$Kur = \frac{C(B(a)-\chi)^4}{\delta^4} \qquad (12)$$

Where, $C(\ )$ represents the expected values of the signal samples.

Equation (13) used to calculate **skewness** of Grayscale statistic features.

$$S(i,a) = \frac{1}{A_s}\sum_m \sum_{n \in q}[u(m-1, n-r) - M(n,q)] \qquad (13)$$

Where, $u$ represents Inverse Difference Normalized image, $m$ represents Informative Measure of Correlations and $q$ indicates spatial relationship of pixel. Haralick texture features likes contrast, energy, entropy, homogeneity are discussed below,

**Contrast** can be expressed in equation (14)

$$Contrast = \sum_{x=0}^{m-1}\sum_{y=0}^{m-1}(x-y)^2 A(x,y) \qquad (14)$$

Here, $m$ is total number of gray levels in preprocessed output image, $A$ is ACEWT matrix. $x$ and $y$ denotes mean value of $A_i$ and $A_j$ respectively.

**Energy** can be expressed in equation (15)

$$Energy = \sum_{x=0}^{m-1}\sum_{y=0}^{m-1} A(x,y)^2 \qquad (15)$$

**Entropy** can be expressed in equation (16)

$$Entro = -\sum_{x=0}^{2m-2} A_{i+j}(x)\log(A_{i+j}(j)) \qquad (16)$$

**Homogeneity** can be expressed in equation (17)

$$Homogeneity = \sum_{x=0}^{m-1}\sum_{y=0}^{m-1} \frac{A(x,y)}{1+|x-y|} \qquad (17)$$

Thus the above features are extracted by ACEWT. Then these extracted features are given into DIEZNN for classification.

### 3.4 Classification using Double Integral Enhanced Zeroing Neural Network

Classification of CT images using DIEZNN for lung cancer classification [14] is discussed. The extracted features given to classification stage. Here, CT image is classified into cancer and non-cancer. The IEZNN model has some noise suppression capabilities when classifying CT images; however it struggles through linear noises. Consequently, a new model required to account for presence of linear noise. So DIEZNN model is proposed in this paper. The error function $T(s)$ and classification function $\dot{T}(s)$ is related in equation (18)

$$\dot{T}(s) + T(s) = 0 \qquad (18)$$

The unique ZNN design formula through double integral item inferred for solving classification problem of time-varying matrix under linear disturbances, which is given in equation (19)

$$K(s) = -\mu \int_0^s K(\rho) \qquad (19)$$

Where, $\mu$ indicates the linear noises present in the image. $K(\rho)$ denotes the extracted features. $s$ refers a time interval. A unique ZNN model is first built based on the double integral's design formula. The derivative of $H(s)$ value is given in equation (20)

$$H(s) = -\mu \int_0^s H(\rho)d\rho \qquad (20)$$

Where, $H(\rho)$ denotes required features. The DIEZNN model's design equation is given in equation (21)

$$\dot{T}(s) = -(2\mu+1)T(s) - (\mu^2 + 2\mu)\int_0^s T(\rho) - \mu^2 \int_0^s \int_0^\rho T(\beta)d\beta d\rho \qquad (21)$$

Where, $\beta$ is a random value. Noise must be considered because it is unavoidable in many practical industrial domains. The final classification model equation is given in equation (22)

$$\dot{T}(s) = -\eta T(s) - \varphi \int_0^s T(\rho)d\rho - \mu^2 \int_0^s \int_0^\rho T(\beta)d\beta d\rho + G(s) \qquad (22)$$

Where $G(s)$ indicates matrix-form noise value. $\varphi$ is a weight parameter of DIEZNN and $\eta$ is a random value. Thus the DIEZNN classifies the input CT image into cancer and non-cancer.

Because of its convenience, pertinence, the artificial intelligence-based optimization strategy is taken into account in the DIEZNN classifier. In this work, ALSOA is employed to optimize the DIEZNN optimum parameter $\varphi$. Here, ALSOA is employed for tuning the weight and bias parameter of DIEZNN.

### 3.5 Optimization using Artificial Lizard Search Optimization Algorithm

Here, $\varphi$ parameter of DIEZNN is optimized by Artificial Lizard Search Optimization Algorithm [15]. The proposed ALSOA displays its numerical modeling for resolving optimization issues. The redheaded Agama lizards prepared for foraging respective random locations (multi-dimensional search space) are taken into account by the artificial lizard search optimization.

### 3.5.1 Stepwise procedure of Artificial Lizard Search Optimization Algorithm

The recommended ALSOA, Lizards hunt for food through leaping from location to prey in order to get the food (insects). During their leap, lizards vary their angular position and velocity. Here is a stepwise procedure of the ALSOA.

**Step 1: Initialization**

The initialization of the variables proceeds the ALSO Algorithm. In a multidimensional search space, the location of $m$ jumping lizards can be represented by a vector, as given in equation (23)

$$KK = \begin{bmatrix} KK_{1,1} & KK_{1,2} & ... & KK_{1,w} \\ KK_{2,1} & KK_{2,2} & ... & KK_{2,w} \\ ... & ... & ... & ... \\ KK_{m,1} & KK_{m,2} & ... & KK_{m,w} \end{bmatrix} \quad (23)$$

Where, $KK_{u,v}$ indicates the $u^{th}$ jumping lizard's $v^{th}$ dimension.

**Step 2: Random generation**

Afterward the initialization generated the input parameters stochastically. The ideal fitness are selected depends upon its obvious hyper parameter condition.

**Step 3: Fitness Function Estimation**

The classification accuracy and the amount of chosen features are taken into account by the fitness function. To evaluate individual solutions, the following fitness function is utilized, which is given in equation (24)

$$Fitness\ fn = optimizing\ \phi \qquad (24)$$

**Step 4: Exploration phase**

The lizard's leap depends in large part on torque. In the following iterations, $\gamma$ is utilized to update derivatives of each lizard's body, tail angles. Here, $e(\beta_u(l))$ denotes the $u^{th}$ lizard's angular location in the $l^{th}$ iteration, respectively, and it is defined in equation (25)

$$e(\beta_u(l)) = \left[\beta_{uy}(l), \frac{\beta_{uy}^2(l) - y}{w - f}, \beta_{uy}(l), \frac{-\beta_{us}^2(l) + z}{w - f}\right] \qquad (25)$$

Where, $\beta_{us}(l)$ and $\beta_{uy}(l)$ indicates the tail angle derivative and body angle derivative. $y$ and $c$ denotes the position vector. $w$, $l$ indicates the lizard's movement value.

**Step 5: Exploitation phase**

Each particle's fitness is assessed, and the $kbest$ value for current iteration updated from its better location. Additionally, the position vector of $u^{th}$ lizard in the $l^{th}$ iteration updated together through updating of angular position, velocity, torque and derivatives of the body, tail angles then which is given in equation (26)

$$a_u(l+1) = a_u(l) + \gamma_u(l+1) \times (0.3) \times \nabla\phi \times r \times (kbest(l) - a_u(l)) \qquad (26)$$

Where, $\nabla\phi$ denotes radian difference among the body, tail angles. $\gamma$ represents the torque value. $a_u(l)$ indicates the $l^{th}$ iteration's position vector. $r$ is random value.

**Step 6: Termination**

The weight parameter value $\phi$ from Artificial Lizard Search Optimization Algorithm are optimized by ALSO, repeat step 3 iteratively until fulfill halting criteria $u = u + 1$. Then LCC-DIEZNN-ALSO-CTI categorizes CT images into cancer, non-cancer with accurately by minimizing the computation period with error.

## 4. Result with discussion

The experimental outcomes of proposed method are discussed in this section. The proposed LCC-DIEZNN-ALSO-CTI approach is implemented in MATLAB using lung cancer dataset. The obtained outcome of the proposed LCC-DIEZNN-ALSO-CTI approach is analyzed with existing systems like LCC-AHHMM-CT [16], LCC-ICNN-CT [17], and LCC-RFCN-MLRPN-CT [18] respectively.

### 4.1 Performance measures

This is a crucial step for choosing the optimal classifier. Performance measures are assessed to assess performance, including accuracy and ROC.

### 4.1.1 Accuracy

Accuracy measures the proportion of samples (positives and negatives) besides total samples and it is given by the equation (27),

$$accuracy = \frac{TP + TN}{TP + TN + FN + FP} \tag{27}$$

Where, $TN$ signifies True Negative, $TP$ signifies True Positive, $FN$ signifies False Negative and $FP$ signifies False Positive.

### 4.1.2 ROC

It is given by the equation (28)

$$ROC = 0.5 \times \left( \frac{TP}{TP + FN} + \frac{TN}{TN + FP} \right) \tag{28}$$

### 4.2 Performance Analysis

Fig 2-3 depicts simulation results of proposed LCC-DIEZNN-ALSO-CTI method. Then, the proposed LCC-DIEZNN-ALSO-CTI method is likened with existing LCC-AHHMM-CT, LCC-ICNN-CT, and LCC-RFCN-MLRPN-CT methods respectively.

Figure 2 displays Accuracy analysis. Here, LCC-DIEZNN-ALSO-CTI attains 18.32%, 27.20%, and 34.32% higher accuracy for cancer; 22.80%, 33.91% and 19.75% greater accuracy for non-cancer analyzed with existing LCC-AHHMM-CT, LCC-ICNN-CT, and LCC-RFCN-MLRPN-CT methods.

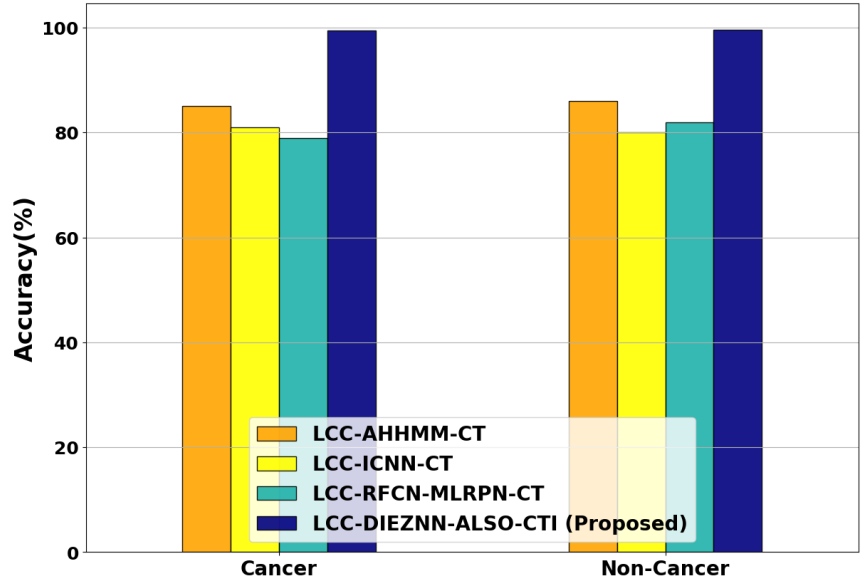

**Figure2: Performance Analysis of Accuracy**

Figure 3 displays RoC analysis. Here, LCC-DIEZNN-ALSO-CTI attains 20.94%, 22.46%, and 30.15% higher RoC for cancer; 25.35%, 21.73% and 27.98% higher ROC for non-cancer analyzed with existing LCC-AHHMM-CT, LCC-ICNN-CT, and LCC-RFCN-MLRPN-CT methods.

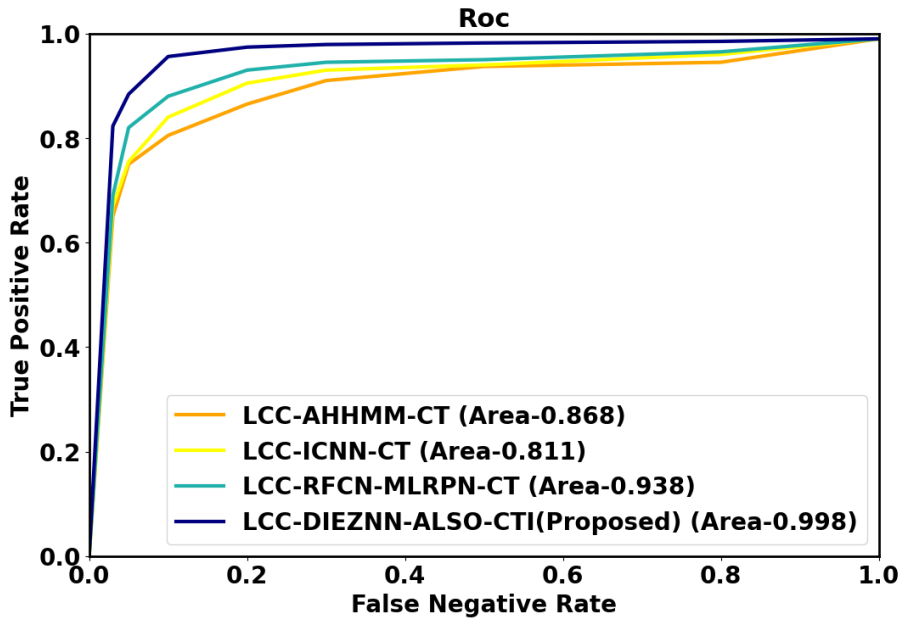

**Figure 3: Performance Analysis of RoC**

## 4.3 Discussion

The primary goals of the current effort were to assess the proposed LCC-DIEZNN-ALSO-CTI's lung cancer classification capabilities as well as their added benefits. DIEZNN model is used for performing the classification process. The proposed method is contrasted with three existing methodologies. When analyzed to traditional deep learning processes, the proposed ADCN-FAGAN-POA approach achieves the highest detection accuracy. The proposed method has better RoC score of 0.9971 that is 2.7% to 5.8% higher than other methods. The outcomes show that proposed approach is effective at classification of lung cancer CT image.

## 5. Conclusion

In this section, Double Integral Enhanced Zeroing Neural Network Optimized with ALSOA fostered Lung Cancer Classification utilizing CT Images (LCC-DIEZNN-ALSO-CTI) is successfully implemented. The proposed LCC-DIEZNN-ALSO-CTI approach is implemented in MATLAB utilizing lung cancer dataset. The performance of the proposed LCC-DIEZNN-ALSO-CTI approach attains 20.94%, 22.46% and 30.15% higher RoC analyzed with existing LCC-AHHMM-CT, LCC-ICNN-CT, and LCC-RFCN-MLRPN-CT methods respectively.